\documentclass{emulateapj}
\begin{document}

\title{UV properties of early-type galaxies in the Virgo cluster.}
\author{
A. Boselli\altaffilmark{1}, L. Cortese\altaffilmark{1}, 
J.M. Deharveng\altaffilmark{1},
G. Gavazzi\altaffilmark{2}, 
K. S. Yi \altaffilmark{3},
A. Gil de Paz\altaffilmark{4}, 
M. Seibert\altaffilmark{5},  S. Boissier \altaffilmark{4},J. Donas\altaffilmark{1},
%,T. Barlow\altaffilmark{5}, L. Bianchi\altaffilmark{6},
%Y.-I. Byun\altaffilmark{7},
%K. Forster\altaffilmark{5}, P. G. Friedman\altaffilmark{5},
%T. M. Heckman\altaffilmark{8}, 
%P. Jelinsky\altaffilmark{9},
Y.-W. Lee\altaffilmark{6}, B. F. Madore\altaffilmark{4,7},
%R. Malina\altaffilmark{2},
D. C. Martin\altaffilmark{5}, 
%B. Milliard\altaffilmark{2},
%P. Morrissey\altaffilmark{5}, S. Neff\altaffilmark{11},
R. M. Rich\altaffilmark{8}, 
Y.-J. Sohn\altaffilmark{6}
%D. Schiminovich\altaffilmark{5},
%O. Siegmund\altaffilmark{9}, 
%T. Small\altaffilmark{5},
%A. S. Szalay\altaffilmark{8}, 
%B. Welsh\altaffilmark{9}, T. K. Wyder\altaffilmark{5}
}
\altaffiltext{1}{Laboratoire d'Astrophysique de Marseille, BP8, Traverse du Siphon, F-13376 Marseille, France}
\altaffiltext{2}{Universit\`a degli Studi di Milano - Bicocca, P.zza della Scienza 3, 20126 Milano, Italy}
\altaffiltext{3}{Oxford University, Astrophysics, Oxford OX1 3RH, United Kingdom}
\altaffiltext{4}{Observatories of the Carnegie Institution of Washington,
813 Santa Barbara St., Pasadena, CA 91101}
\altaffiltext{5}{California Institute of Technology, MC 405-47, 1200 East California Boulevard, Pasadena, CA 91125}
%\altaffiltext{6}{Center for Astrophysical Sciences, The Johns Hopkins University, 3400 N. Charles St., Baltimore, MD 21218}
\altaffiltext{6}{Center for Space Astrophysics, Yonsei University, Seoul 120-749, Korea}
%\altaffiltext{8}{Department of Physics and Astronomy, The Johns Hopkins University, Homewood Campus, Baltimore, MD 21218}
%\altaffiltext{9}{Space Sciences Laboratory, University of California at Berkeley, 601 Campbell Hall, Berkeley, CA 94720}
\altaffiltext{7}{NASA/IPAC Extragalactic Database, California Institute of Technology, Mail Code 100-22, 770 S. Wilson Ave., Pasadena, CA 91125}
%\altaffiltext{11}{Laboratory for Astronomy and Solar Physics, NASA Goddard Space Flight Center, Greenbelt, MD 20771}
\altaffiltext{8}{Department of Physics and Astronomy, University of California, Los Angeles, CA 90095}

\begin{abstract}

We study the UV properties of a volume limited sample of early-type galaxies in the Virgo cluster
combining new GALEX far- (1530 \AA) and near-ultraviolet (2310 \AA) data with spectro-photometric data
available at other wavelengths. The sample includes 264 ellipticals, lenticulars and dwarfs
spanning a large range in luminosity ($M_B$ $\leq$ -15).
While the NUV to optical or near-IR color magnitude relations (CMR)
are similar to those observed at optical wavelengths, with a monotonic reddening of 
the color index with increasing
luminosity, the $(FUV-V)$ and $(FUV-H)$ CMRs show a discontinuity between massive and dwarf objects. 
An even more pronounced dichotomy is observed in the $(FUV-NUV)$ CMR.
For ellipticals the $(FUV-NUV)$ color
becomes bluer with increasing luminosity and with increasing reddening of
the optical or near-IR color indices. For the dwarfs the opposite trend is observed.
These observational evidences are consistent with the idea that the
UV emission is dominated by hot, evolved stars in giant systems, 
while in dwarf ellipticals residual star formation activity is more common.

\end{abstract}


\section{Introduction}

\setcounter{footnote}{0}

The excess ultraviolet radiation from early-type galaxies arises 
from hot stars in late stages of stellar evolution (O'Connell 1999). 
Whether the so-called UV-upturn depends on the detailed galaxy morphology (ellipticals vs. lenticulars) 
and, among ellipticals, on luminosity (dEs vs. giant Es) is yet unknown.
It would not be surprising if the UV properties of dwarf elliptical galaxies   
would differ from those of giants, given that other structural (Gavazzi et al. 2005) and 
kinematic (Van Zee et al. 2004) properties depend on luminosity, 
due to different star formation histories (single episodic vs. burst)
(Ferguson \& Binggeli 1994; Grebel 1999).\\
%Beside known structural (Gavazzi et al. 2005) and kinematic (Van Zee et al. 2004) dependencies on luminosity,
%systematic differences in the UV properties of giant and dwarf elliptical galaxies   
%might arise from their different star formation histories (single burst vs. episodic)
%(Ferguson \& Binggeli 1994; Grebel 1999).
%\\

Due to morphological segregation (Whitmore et al$.$ 1993), nearby clusters are the ideal
targets for assembling complete, volume limited samples of early-type objects. 
As part of a study aimed at analyzing the environmental dependence of galaxy evolution, 
we observed large portions of the Virgo cluster with GALEX (Boselli et al$.$ 2005). 
Owing to the superior quality of the photographic material obtained by
Sandage and collaborators, an extremely accurate and homogeneous morphological classification
exists for Virgo galaxies, down to $m_B$ $\leq$ 18 mag ($M_B$ $\leq$-13 assuming a distance of 17 Mpc),
allowing a detailed discrimination among different subclasses of early-type
galaxies (ellipticals, lenticulars, dwarfs) and an exclusion of 
quiescent spirals. Furthermore a wealth of ancillary data for many Virgo members, 
covering a large portion of the
electromagnetic spectrum from the visible to the infrared is available from  
the GOLDMine database (Gavazzi et al$.$ 2003). 
%The new GALEX data of the Virgo cluster thus provide us with a unique opportunity for studying, for the first 
%time, the UV properties of dwarf ellipticals and lenticulars, and to compare them to those of bright
%ellipticals using a statistically significant sample.}

\section{Data}

The analysis presented in this work is based on an optically selected sample of 
early-type galaxies including 
giant and dwarf systems (E, S0, S0a, dE and dS0) extracted from the Virgo 
Cluster Catalogue of Binggeli et al. (1985), 
which is complete to $m_B$ $\leq$18 mag ($M_B$ $\leq$ -13). The Virgo cluster region 
was observed in spring 2004 
as part of the All Imaging Survey (AIS) and of the Nearby Galaxy Survey (NGS) carried out by the 
Galaxy Evolution Explorer (GALEX) 
in two UV bands: FUV ($\rm \lambda_{eff}=1530\AA, \Delta \lambda=400\AA$) 
and NUV ($\rm \lambda_{eff}=2310\AA, \Delta \lambda=1000\AA$), covering 427 objects.
Details of the GALEX instrument and data characteristics can be found in 
Martin et al$.$ (2005) and Morrissey et al$.$ (2005).\\

The present sample includes all Virgo cluster early-type systems
detected in the NUV GALEX band (264 objects, 194 from the NGS); of these, 126 (of which 74 from the 
NGS) have been also detected in the FUV. 
The resulting sample is thus ideal for the proposed analysis
as it provides us with the first large volume-limited sample 
of elliptical, lenticular and dwarf galaxies spanning 
4 dex in luminosity with homogeneous data.
Whenever available, we extracted fluxes from the deep NGS images, obtained with 
an average integration time
of $\sim$ 1500 sec, complete to $m_{AB}$ $\sim$ 21.5 in the NUV and FUV. 
Elsewhere UV fluxes have been extracted from the less deep 
AIS images ($\sim$ 70 sq. degrees), obtained with an average integration time of $\sim$ 100 sec, 
complete to $m_{AB}$ $\sim$ 20 in both the FUV and NUV bands. 
The resulting sample, although not complete in both UV bands,
includes giants and dwarf systems: at a limiting magnitude of $M_B$ $\leq$ -15, 71 \% 
of the observed galaxies have been detected in the NUV, 46\% in the FUV.

All UV images come from the GALEX IR1.0 release. 
UV fluxes were obtained by integrating GALEX images
within elliptical annuli of increasing diameter up to the optical B band 25 mag 
arcsec$^{-2}$ isophotal radii consistently
with the optical and near-IR images. Independent measurements of the same 
galaxies obtained in different 
exposures give consistent photometric results within 10\% in the NUV and 15\% 
in the FUV in the AIS, and  about
a factor of two better for bright (NUV $\leq$16) galaxies. 
The statistical uncertainty in the UV photometry 
is on average a factor of $\sim$ 2 better in the NGS than in the AIS especially
for fainter objects.\\

UV data have been combined with multifrequency data taken from the 
GOLDMine database (http:\slash \slash goldmine.mib.infn.it; Gavazzi et al$.$ 2003). 
These are B and V imaging data, mostly
from Gavazzi et al$.$ (2005) and Boselli et al$.$ (2003), and near-IR H imaging 
from Gavazzi et al$.$ (2000, 2001). Optical and near-IR data
have on average a photometric precision of $\sim$ 10\%. Spectroscopic 
metallicity index Mg$_2$ and velocity dispersion data come from GOLDMine or 
from Golev \& Prugniel (1998) and Bernardi et al$.$ (2002).
Unless specified, we adopt the homogeneous morphological classification of Binggeli et al$.$ (1985) 
based on high quality photographic material.  \\

Galaxies analyzed in this work are all bona-fide Virgo cluster members: 
given the 3-D structure of the cluster, distances 
have been assigned following the subcluster membership criteria of Gavazzi et al$.$
(1999). Owing to the high galactic latitude of Virgo, 
no galactic extinction correction was applied ($A_B$ $\leq$ 0.05).

\section{The UV properties of early-type galaxies}

Despite the complex 3-D structure of Virgo (Gavazzi et al$.$ 1999), 
the uncertainty on the distance (hence on the luminosity) 
of the target galaxies, does not constitute a major source of dispersion in the
determination of the color-magnitude (CMR) relation. 
Figure 1 shows various UV to optical and
near-IR CMRs. Similar results are obtained if, instead of the mass-tracer H band luminosity
(Zibetti et al$.$ 2004), we use the B band absolute magnitude.

\begin{figure*}
\epsscale{0.85}
\plotone{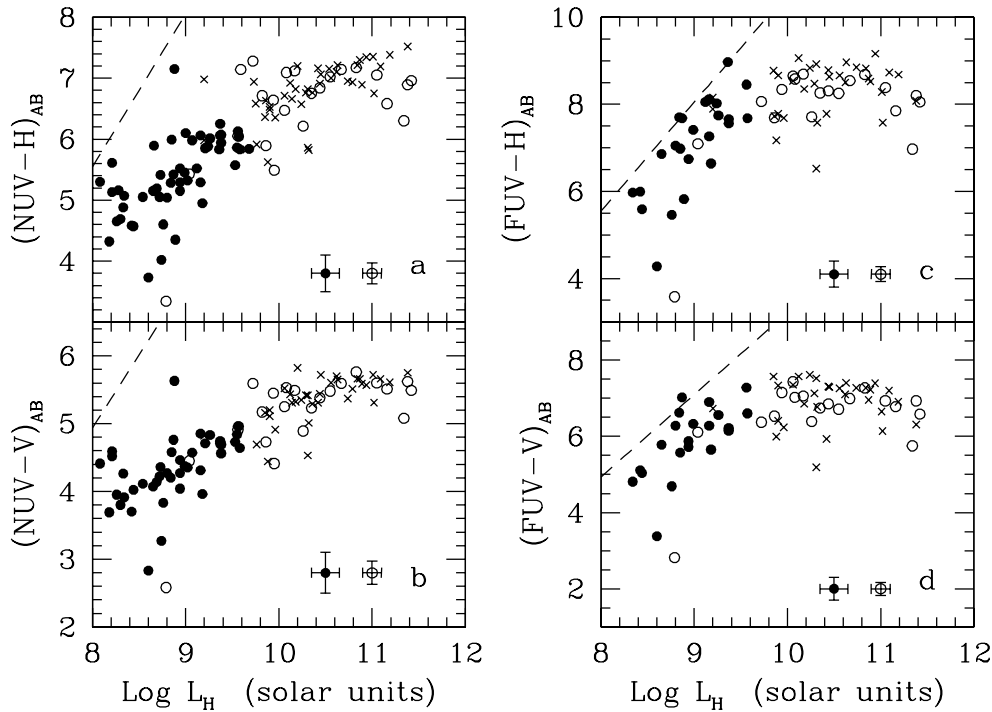}
\small{\caption{The near-UV (left column) and far-UV (right column) to optical and near-IR 
color magnitude relations. Colors are in the AB magnitude system. Open circles are for ellipticals, 
filled circles for dwarfs, crosses for lenticulars (S0-S0a). 
Galaxies redder than the dashed line are undetectable
by the present survey (at the NGS limit). Largest 1$\sigma$ errors for luminous and dwarf systems are given.}}
\label{region}
\end{figure*}

%\noindent
The NUV to optical (Fig$.$~1b) and near-IR (Fig$.$~1a) CMRs are well defined 
and are similar to optical or near-IR
CMRs, with brighter galaxies having redder colors, independent of their morphological type: 
the color index $(NUV-V)$ increases
by $\sim$ 2 magnitudes from dwarfs ($L_H$  $\sim$  $10^8$ L$_{H \odot}$)
to giants ($L_H$ $\sim$  $10^{11.5}$ L$_{H \odot}$), while $(NUV-H)$ changes by $\sim$ 3 mag.
A weak flattening of the relation appears for $L_H$ $\geq$ 10$^{10}$ L$_{H \odot}$. 
This behavior confirms the one reported by Ferguson (1994) in the $(B-V)$ vs. $M_B$ CMR.\\

On the contrary, the FUV to optical (Fig$.$~1d) and near-IR (Fig$.$~1c) CMRs 
differ systematically for dwarfs and giant
systems: galaxies brighter than $L_H$ $\sim$ $10^{9.5}$ L$_{H \odot}$ 
have similar red colors, while for $L_H$ $\leq$ $10^{9.5}$ L$_{H \odot}$ colors become progressively
bluer. 
Even if this trend can be due to a selection effect, 
(reddest dwarfs being undetectable in the FUV), it is indisputable that there exists 
a significant population of dEs with bluer colors than Es and S0s. 
%In both NUV and FUV CMRs the dispersion increases
%at lower luminosities, and seems larger in lenticulars (crosses) than in ellipticals 
%(open circles; {\bf see Table 1}). 
%Notice however that 
%the lack of red, low-luminosity galaxies can be due to a selection effect, 
%the reddest dwarfs being undetectable in the UV. 

A similar pattern (systematic differences between massive and dwarf systems and
between NUV and FUV to optical or near-IR color indices) is obtained using 
color-color diagrams (not shown).

\begin{table}
\label{tablefit}
\caption {Main relations for early type galaxies}
\[
\begin{array}{p{0.09\linewidth}ccccc}
\hline
\noalign{\smallskip}
x  & y  & a & b & R  & rms \\
\noalign{\smallskip}
\hline
\noalign{\smallskip}									       
 &  &  {Ellipticals^1}&   \\
\noalign{\smallskip}
\hline
\noalign{\smallskip}									       
$L_H$	&   FUV-NUV  & -0.30\pm0.14 &  + 4.52\pm1.52     &-0.47 & 0.31  \\
$L_H$	&   FUV-H    & -0.22\pm0.19 &  + 10.55\pm2.10    &-0.28 & 0.43  \\
$L_H$	&   NUV-H    & 0.17\pm0.18 &   + 4.85\pm1.85     &0.22  & 0.47  \\
$L_H$	&   FUV-V    & -0.15\pm0.18 &   + 8.38\pm1.88    &-0.21 & 0.38  \\
$L_H$	&   NUV-V    & 0.26\pm0.12 &   + 2.55\pm1.30     &0.45  & 0.31  \\
$B-H$	&   FUV-NUV  & -0.84\pm0.45 &   + 3.22\pm0.98    &-0.43 & 0.32  \\
%Mg2	&   FUV-NUV  &     &      &      \\
$\sigma$&   FUV-NUV  &  -1.35\pm0.37 &   + 4.39\pm0.89    &-0.69 & 0.26 \\
\hline
\noalign{\smallskip}
 &  &  {Lenticulars}&   \\
\noalign{\smallskip}
\hline
\noalign{\smallskip}									       
$L_H$	&   FUV-NUV  & -0.28\pm0.15 &  + 4.40\pm1.62        &  -0.31& 0.45  \\
$L_H$	&   FUV-H    & 0.31\pm0.21 &  + 0.75\pm2.00         &  0.27 & 0.58  \\
$L_H$	&   NUV-H    & 0.61\pm0.11 &  + 0.51\pm1.17         &  0.65 & 0.36  \\
$L_H$	&   FUV-V    & 0.03\pm0.23 &   + 6.62\pm2.38        &  0.03 & 0.59  \\
$L_H$	&   NUV-V    & 0.49\pm0.09 &   + 0.26\pm1.00        &  0.68 & 0.25  \\
$B-H$	&   FUV-NUV  & -1.00\pm0.32 &   + 3.70\pm0.70       &  -0.49& 0.42  \\
%Mg2	&   FUV-NUV  &         &       &      \\
$\sigma$&   FUV-NUV  & -1.29\pm0.39 &   + 4.28\pm0.84       &  -0.58& 0.39  \\
\hline
\noalign{\smallskip}
 &  &  {Dwarfs}&   \\
\noalign{\smallskip}
\hline
\noalign{\smallskip}									       
$L_H$	&   FUV-NUV  & 1.73\pm0.41 &  - 13.90\pm2.16       &  0.52 & 0.59 \\
$L_H$	&   FUV-H    & 2.55^*\pm0.55 &  - 15.97^*\pm4.96       &  0.68^* & 0.91^* \\
$L_H$	&   NUV-H    & 0.91\pm0.19 &  - 2.72 \pm1.68       &  0.56 & 0.57 \\
$L_H$	&   FUV-V    & 1.91^*\pm0.55 &   + 11.35^*\pm4.93      &  0.60^* & 0.87^* \\
$L_H$	&   NUV-V    & 0.63\pm0.17 &   + 1.28\pm1.05       &  0.49 & 0.47 \\
$B-H$	&   FUV-NUV  & 0.95\pm0.45 &   + 0.12\pm0.73       &  0.40 & 0.60 \\
%Mg2	&   FUV-NUV  &        &       &     \\
$\sigma$&   FUV-NUV  &    -    &  -    &  -   \\

\noalign{\smallskip}
\hline
\end{array}
\]
Notes to Table:
Col. 1 and 2: $x$ and $y$ variables;
Col. 3 and 4: slope $a$ and intercept $b$ of the bisector linear fit with weighted variables;
Col. 5: Pearson correlation coefficient;
Col. 6: mean dispersion around the best fit;
1: excluding VCC 1499;
*: uncertain values because of the UV detection limit
\end{table}

\begin{figure*}
\epsscale{0.7}
\plotone{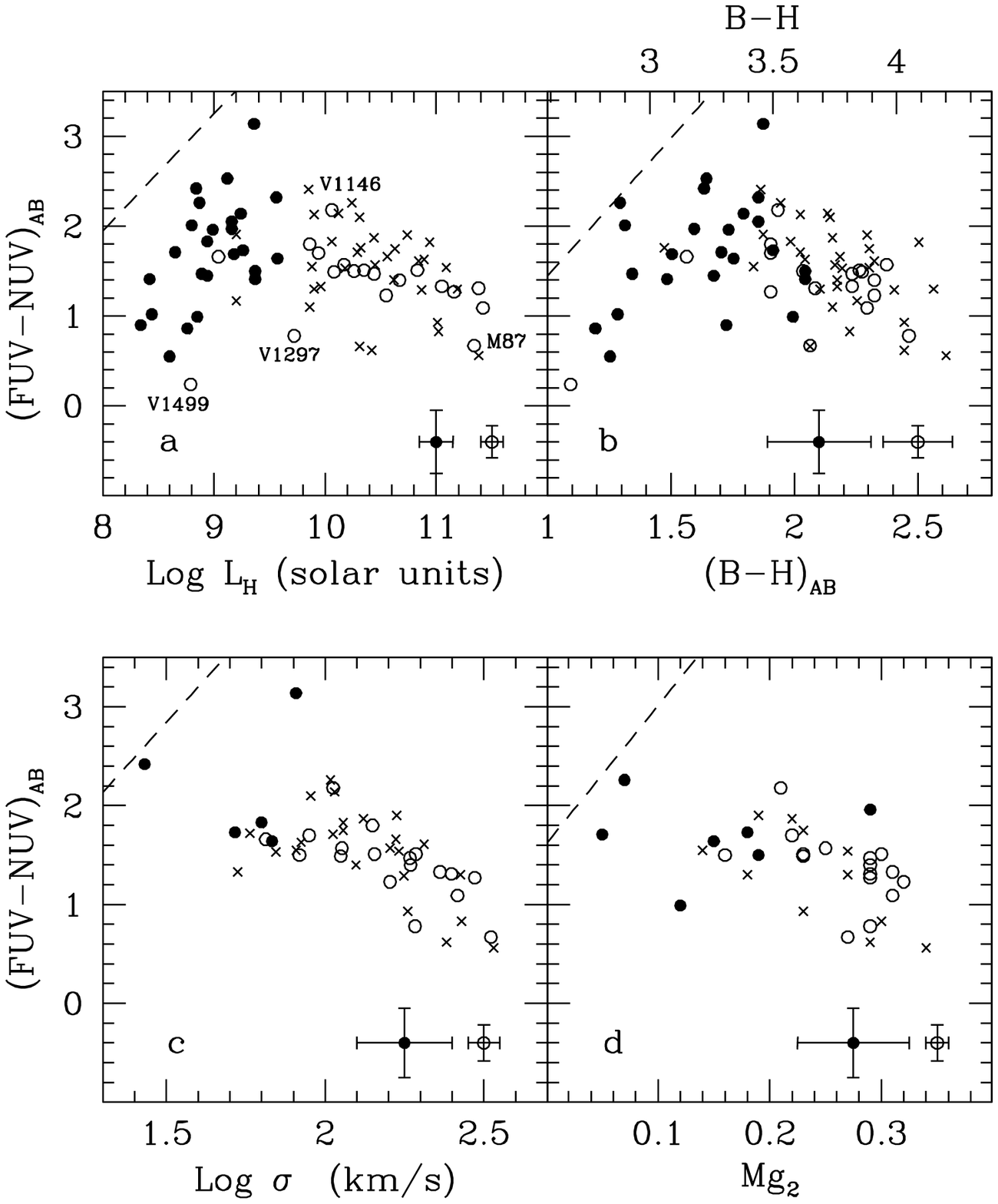}
\small{\caption{The relationship between the UV color index $(FUV-NUV)$ and a) the total H band luminosity,
b) the B-H color index, 
c) the logarithm of the central velocity dispersion and d) the Mg$_2$ index. 
Symbols are as in Fig. 1.
Labeled points indicate objects having unusual radio or optical properties (see Sect. 3).}}
\label{region}
\end{figure*}

The dichotomy between giants and dwarfs is even more apparent in the UV color index 
$(FUV-NUV)$ (see Fig. 2).
The $(FUV-NUV)$ becomes redder with increasing luminosity for dwarf ellipticals while, on the contrary, 
it becomes bluer for giant ellipticals (Fig. 2a). 
The blueing relation is tight among ellipticals (see Table 1) and barely observed 
in lenticulars because of their higher dispersion
\footnote{The scatter in the blueing relation among ellipticals decreases significantly 
(from 0.31 to 0.10)
if we exclude the misclassified post-starburst 
dwarf VCC 1499 (Gavazzi et al$.$ 2001; Deharveng et al$.$ 2002), 
the radio galaxy M87, VCC 1297 (the highest surface brightness galaxy
in the sample of Gavazzi et al$.$ (2005)) and VCC 1146.  
Beside its extremely high 
surface brightness, making VCC 1297 a non standard object, we do not have any evidence 
indicating a peculiar star formation history or present nuclear activity in VCC 1297 and VCC 1146 
that could justify their exclusion.}.\\
A similar behavior between ellipticals and lenticulars 
is observed in the $(FUV-NUV)$ color relation (Fig. 2b): this mixed giant population becomes bluer in the UV 
with increasing reddening in the $(B-H)$ color index. \\
The behavior of dwarf ellipticals is different: although with a huge dispersion, 
the $(FUV-NUV)$ color index reddens as the $(B-H)$, $(B-V)$ and $(FUV-V)$ indices (the two last not shown).\\
The dichotomy between dwarf and giant systems cannot be observed in the run of $(FUV-NUV)$ vs. 
central velocity dispersion (which is directly related to the system total dynamical 
mass; Fig. 2c) nor as a function of the 
metallicity sensitive (Poggianti et al$.$ 2001) Mg$_2$ Lick index (Fig. 2d) 
because these two parameters are not available for dwarfs.
In ellipticals and lenticulars the UV color index $(FUV-NUV)$ depends on both 
the metallicity index Mg$_2$ and $\sigma$ 
in a way opposite to the behavior at optical wavelengths, 
where galaxies are redder when having higher Mg$_2$ and velocity dispersions.

\section{Discussion and conclusion}

For the first time the UV properties of early-type galaxies have been studied 
down to $M_B$ $\sim$ -15 mag. The comparison with previous studies is thus limited to 
the brightest objects.
Our CMR can be compared with the one obtained by Yi et al$.$ (2005) 
based on a complete sample of bright early-type objects ($M_r$ $\leq$ -20 mag) 
extracted from the Sloan Digital Sky Survey
(SDSS) by Bernardi et al$.$ (2003). The CMR presented by Yi et al$.$ (2005, $NUV-r$ vs. 
$M_r$) shows a significantly larger dispersion ($\sigma \geq$ 1.5 mag) than the one found in Virgo 
(see Table 1). 
As discussed in Yi et al$.$
(2005), the large dispersion in their CMR can be ascribed to galaxies with a
mild or residual star formation activity included in the Bernardi et al$.$ (2003)
sample. If restricted to the "UV weak" sample, the dispersion in the Yi et al$.$ relation drops to 0.58
mag, i.e. still larger than the one seen in the Virgo cluster in the same luminosity range. 
Despite possible larger distance uncertainties in the
SDSS, the difference in the scatter between our and the Yi et al$.$ (2005) CMR
might arise from the classification in the SDSS 
that uses concentration indices and luminosity profiles
in discriminating hot from rotating systems.
It is in fact conceivable that the larger dispersion in the CMR of "UV weak"
galaxies of Yi et al$.$ (2005) comes from the contamination of quiescent, 
bulge-dominated Sa spiral disks, that have
structural (concentration indices and light profiles) or population properties (colors and spectra) 
similar to ellipticals and lenticulars (Scodeggio et al$.$ 2002; Gavazzi et al$.$ 2002).\\

The monotonic increase of the $(NUV-V)$ and $(NUV-H)$ colors with luminosity, 
similar to the one observed in the
visible bands by Ferguson (1994) and Ferguson \& Binggeli (1994) 
strongly suggests that  both in dwarfs and giant systems the NUV 2310 \AA~ flux is  
dominated by the same stellar population emitting at longer wavelengths. 
The plateau and the higher dispersion
observed in the FUV CMRs confirm that the UV upturn is observable only 
in the FUV GALEX band at 1530 \AA, as already remarked 
by Dorman et al$.$ (2003) and Rich et al$.$ (2005). \\
%The $(FUV-NUV)$ color index seems therefore to be the ideal
%tracer of the underlying stellar population responsible for UV upturn in early-type galaxies.\\

%%%QUESTO PRIMA DELLE MODIFICHE CHIESTE DALLO SMERIGLIA MARONS 
%The observed trend between $(FUV-NUV)$
%and the metallicity sensitive Mg$_2$ index, 
%reproduced by models (Bressan et al$.$ 1994; Yi et al$.$ 1998),
%confirms the early IUE result of Burstein et al$.$ (1988).
%Conversely Rich et al$.$ (2005) did not find any correlation between
%the color index $(FUV-r)$ and Mg$_2$ nor with the velocity dispersion $\sigma$
%in a large sample of SDSS early-type galaxies observed by GALEX.
%Their lack of correlation might derive from insufficient 
%dynamic range in Log $\sigma$ (2.1-2.4 km s$^{-1}$) and Mg$_2$ (0.18-0.30).\\

The mild trend between the $(FUV-NUV)$ color and the metallicity
sensitive Mg$_2$ index, as seen in the present dataset, is predicted
by models (Bressan et al$.$ 1994; Yi et al$.$ 1998). It is also in
qualitative agreement with the early IUE result of Burstein et al$.$
(1988). However, it should be noted that the Burstain dataset is
systematically different from the GALEX dataset: Burstain et al. used
a different color index, $(1550\AA-V)$, which was computed inside a
fixed IUE aperture. On the other hand, the trend noted here contrasts
with the analysis of Rich et al. (2005) who recently reported the lack
of a significant correlation between the $(FUV-r)$ color index and
Mg$_2$, and the velocity dispersion $\sigma$. They examined a large
sample of SDSS early-type galaxies, also observed by GALEX. We can 
only speculate that the lack of correlation in the dataset may be the
result of a relatively restricted dynamic range in the variables 
available to those authors at that time: only 2.1-2.4 km s$^{-1}$ in
Log $\sigma$ and 0.18-0.30 in Mg$_2$. Clearly the issue is not decided
and would benefit from targetted new observations.\\

The newest result of the present paper, shown in Fig. 2, addresses the
question raised by O'Connell (1999) concerning the dependence of the UV properties on galaxy morphology.
We have shown that a dichotomy exists between 
giant and dwarf ellipticals and, to a lesser extent, between ellipticals and lenticulars.
The blueing of the UV color index with luminosity,  
metallicity and velocity dispersion indicates that the UV upturn is more important 
in massive, metal rich systems. 

The accurate morphological classification in our sample allow us to discriminate between E and SOs. 
The higher dispersion in the $(FUV-NUV)$ vs. $L_H$ relation 
observed for the lenticulars compared to the extremely tight one for ellipticals (see Table 1), bears
witness to recent, minor episodes of star formations combined with an old stellar 
population, as determined also from kinematic and spectroscopic observations (Dressler \& Sandage 1983; 
Neistein et al$.$ 1999; Hinz et al$.$ 2003). We have shown that the UV properties of
ellipticals are different than those of lenticulars, suggesting a different evolution.  \\

The opposite behavior (reddening of the UV color index with luminosity) of dwarfs with respect to giants, 
similar to that observed for spirals, indicates that the UV spectra of 
low luminosity objects are shaped by the contribution of young stars, 
thus are more sensitive to the galaxy's star formation history than to the metallicity.
Spectroscopy was recently obtained (unpublished) for the 7 bluest dEs ($(FUV-NUV)$ $\leq$ 1.4 mag) in our sample.
Three out of seven show Balmer emission lines, another three strong H$\delta$ in absorption 
(H$\delta$EW$\geq$ 5 \AA) witnessing a present or recent star formation activity.
This implies that the stellar population of dwarfs has 
been formed in discrete and relatively recent episodes,
as observed in other nearby objects (Grebel 1999). \\

More evidences are building up that mass drives the star formation history
in hot systems (Trager et al$.$ 2000; Gavazzi et al$.$ 2002; Caldwell et al$.$ 2003; Poggianti 2004) 
as in rotating ones (Gavazzi et al$.$ 1996, 2002; Boselli et al$.$ 2001) and that 
the stellar population of massive
ellipticals is on average older than that of dwarfs.

\acknowledgements
We thank an unknown referee for his/her criticism.
GALEX (Galaxy Evolution Explorer) is a NASA Small Explorer, launched in April 2003.
We gratefully acknowledge NASA's support for construction, operation,
and science analysis for the GALEX mission,
developed in cooperation with the Centre National d'Etudes Spatiales
of France and the Korean Ministry of Science and Technology. 
The authors would like to take this opportunity to thank the members
of the GALEX SODA Team for their valiant efforts in the timely
reduction of the complex observational dataset covering the full
expanse of the Virgo cluster.

\references

\reference{}Bernardi, M., et al$.$, 2002, AJ, 123, 2990

\reference{}Bernardi, M., et al$.$, 2003, AJ, 125, 1817

\reference{}Binggeli, B., Sandage, A. \& Tammann, G., 1985, AJ, 90, 1681

%\reference{}Boselli, A., Tuffs, R., Gavazzi, G., Hippelein, H. \& Pierini, D., 1997, A\&AS, 121, 507

\reference{}Boselli, A., Gavazzi, G., Donas, J. \& Scodeggio, M., 2001, AJ, 121, 753

\reference{}Boselli, A., Gavazzi, G. \& Sanvito, G., 2003, A\&A, 402, 37

\reference{}Boselli, A., et al$.$, 2005, ApJ, 623, L13

\reference{}Bressan, A., Chiosi, C., \& Fagotto, F., 1994, ApJS, 94, 63

\reference{}Burstein, D., Bertola, F., Buson, L., Faber, S. \& Lauer, T., 1988, ApJ, 328, 440

\reference{}Caldwell, N., Rose, J.A. \& Concannon, K.D., 2003, AJ, 125, 2891

\reference{}Deharveng, JM., Boselli, A., Donas, J., 2002, A\&A, 393, 843

\reference{}Dorman, B., O'Connell, R. W. \& Rood, R. T., 2003, ApJ, 591, 878 

\reference{}Dressler, A. \& Sandage, A., 1983, ApJ, 265, 664 

\reference{}Ferguson, H., 1994, in "Dwarf Galaxies", ESO Conference and Workshop Proceedings, ed. G. Meylan \& P. Prugniel, p.475

\reference{}Ferguson, H. \& Binggeli, B., 1994, A\&ARv, 6, 67

\reference{}Gavazzi, G., Pierini, D., \& Boselli, A., 1996, A\&A, 312, 397

\reference{}Gavazzi, G., Boselli, A., Scodeggio, M., Pierini, D., Belsole, E., 1999, MNRAS,
304, 595

\reference{}Gavazzi, G., Franzetti, P., Scodeggio, M., et al$.$, 2000, A\&AS, 142, 65

\reference{}Gavazzi, G., Zibetti, S., Boselli, A., et al., 2001, A\&A, 372, 29

\reference{}Gavazzi, G., Bonfanti, C., Sanvito, C., Boselli, A., Scodeggio, M., 2002, ApJ,
576, 135

\reference{}Gavazzi, G., Boselli, A., Donati, A., Franzetti, P. \& Scodeggio, M., 2003, A\&A, 400, 451

%\reference{}Gavazzi, G., Zaccardo, A., Sanvito, G., Boselli, A. \& Bonfanti, C., 2004, A\&A, 417, 499

\reference{}Gavazzi, G., Donati, A., Cucciati, O., et al., 2005, A\&A, 430, 411

\reference{}Golev, V. \& Prugniel, P., 1998, A\&AS, 132, 255

\reference{}Grebel, E., 1999, in "The stellar content of Local Group galaxies", proceedings of the 192 IAU, 
1998, PASP, Edited by P. Whitelock and R. Cannon, p.17

\reference{}Hinz, J., Rieke, G. \& Caldwell, N., 2003, AJ, 126, 2622

\reference{}Martin, C., et al$.$, 2005, ApJ, 619, L1

\reference{}Morrissey, P., et al$.$, 2005, ApJ, 619, L7

\reference{}Neistein, E., Maoz, D., Rix, H. \& Tonry, J., 1999, AJ, 117, 2666

\reference{}O'Connell, R., 1999, ARA\&A, 37, 603

\reference{}Poggianti, B., Bridges, T., Mobasher, B., et al, 2001, ApJ, 562, 689

\reference{}Poggianti, B., in Clusters of Galaxies: Probes of Cosmological Structure and Galaxy 
Evolution, from the Carnegie Observatories Centennial Symposia.  Edited by J.S. Mulchaey, A. 
Dressler, and A. Oemler, 2004, p. 246

\reference{}Rich, M., Salim, S., Brinchmann, J., et al$.$, 2005, ApJ, 619, L107

\reference{}Scodeggio, M., Gavazzi, G., Franzetti, P., et al.,
2002, A\&A, 384, 812

\reference{}Trager, S. C., Faber, S. M., Worthey \& G, González, J. J., 2000, AJ, 120, 165

\reference{}Yi, S., Demarque, P., \& Oemler, A, 1998, ApJ, 492, 480

\reference{}Yi, S., Yoon, S., Kaviraj, S., et al$.$, 2005, ApJ, 619, L111

\reference{}Van Zee, L., Skillman, E., Haynes, M., 2004, AJ, 128, 121

\reference{}Whitmore, B., Gilmore, D. \& Jones, C., 1993, ApJ, 407, 489

\reference{}Zibetti, S., Gavazzi, G., Scodeggio, M., Franzetti, P., Boselli, A., 2004, ApJ, 579, 261

%\clearpage

\end{document}